\documentclass[draft,ras]{agutex}
 \usepackage[dvips]{graphicx}
 \setkeys{Gin}{draft=false}
\authorrunninghead{NGUYEN ET AL.}

\titlerunninghead{LINK BUDGET AND ORBITAL ANGULAR MOMENTUM}

\begin{document}

%% ------------------------------------------------------------------------ %%
%
%  TITLE
%
%% ------------------------------------------------------------------------ %%

\title{Antenna Gain and Link Budget for Waves Carrying Orbital Angular Momentum (OAM)\\
\normalsize{"This paper is published to AGU Radio Science and is subject to the American Geophysical Union Copyright. The copy of record will be available at AGU Digital Library: http://onlinelibrary.wiley.com/doi/10.1002/2015RS005772/full"}}
%
% e.g., \title{Terrestrial ring current:
% Origin, formation, and decay $\alpha\beta\Gamma\Delta$}
%

%% ------------------------------------------------------------------------ %%
%
%  AUTHORS AND AFFILIATIONS
%
%% ------------------------------------------------------------------------ %%

%Use \author{\altaffilmark{}} and \altaffiltext{}

% \altaffilmark will produce footnote;
% matching \altaffiltext will appear at bottom of page.

\authors{Duy Kevin Nguyen\altaffilmark{1}, Olivier Pascal\altaffilmark{1}, J\'er\^ome Sokoloff\altaffilmark{1}, Alexandre Chabory\altaffilmark{2}, Baptiste Palacin\altaffilmark{3}, and Nicolas Capet\altaffilmark{3}}

\altaffiltext{1}{LAPLACE, Laboratoire Plasma et Conversion d'Energie, Universit\'e de Toulouse UPS, F-31062 Toulouse, France, e-mail: duy.nguyen@laplace.univ-tlse.fr, olivier.pascal@laplace.univ-tlse.fr, jerome.sokoloff@laplace.univ-tlse.fr.}

\altaffiltext{2}{ENAC, TELECOM-EMA, F-31055 Toulouse, France, e-mail: alexandre.chabory@recherche.enac.fr.}

\altaffiltext{3}{CNES, Centre National d'Etudes Spatiales, antenna department, 18 avenue Edouard Belin, Toulouse, France, e-mail: baptiste.palacin@cnes.fr, nicolas.capet@cnes.fr.}

%\altaffiltext{4}{Division of Hydrologic Sciences, Desert Research
%Institute, Reno, Nevada, USA.}

%\altaffiltext{5}{Dipartimento di Idraulica, Trasporti ed
%Infrastrutture Civili, Politecnico di Torino, Turin, Italy.}

%% ------------------------------------------------------------------------ %%
%
%  ABSTRACT
%
%% ------------------------------------------------------------------------ %%

% >> Do NOT include any \begin...\end commands within
% >> the body of the abstract.

\begin{abstract}
This paper addresses the RF link budget of a communication system using unusual waves carrying an orbital angular momentum (OAM) in order to clearly analyse the fundamental changes for telecommunication applications. The study is based on a typical configuration using circular array antennas to transmit and receive OAM waves. For any value of the OAM mode order, an original asymptotic formulation of the link budget is proposed in which equivalent antenna gains and free-space losses appear. The formulations are then validated with the results of a commercial electromagnetic simulation software. By this way, we also show how our formula can help to design a system capable of superimposing several channels on the same bandwidth and the same polarisation, based on the orthogonality of the OAM. Additional losses due to the use of this degree of freedom are notably clearly calculated to quantify the benefit and drawback according to the case.
\end{abstract}

%% ------------------------------------------------------------------------ %%
%
%  BEGIN ARTICLE
%
%% ------------------------------------------------------------------------ %%

% The body of the article must start with a \begin{article} command
%
% \end{article} must follow the references section, before the figures
%  and tables.

\begin{article}

%% ------------------------------------------------------------------------ %%
%
%  TEXT
%
%% ------------------------------------------------------------------------ %%

"This paper is published to AGU Radio Science and is subject to the American Geophysical Union Copyright. The copy of record will be available at AGU Digital Library: http://onlinelibrary.wiley.com/doi/10.1002/2015RS005772/full"

\section{Introduction}

From Maxwell's theory, it is well-known that electromagnetic waves carry energy and both linear and angular momenta \cite{Poynting1909}. The angular momentum has a spin component (SAM), associated with polarization, and an orbital component (OAM), related to the spatial distributions of the field magnitude and phase \cite{jackson_classical_1999}. The mechanical interaction between matter and these two components of the angular momentum has been theoretically and experimentally proven \cite{Beth1936}, \cite{niemiec2012}. 

The SAM with two orthogonal states is well-known and widely exploited in operating systems to double the communication capacity. OAM has not yet been utilized in radio communications, even though it may represent a fundamental new degree of freedom \cite{Djordjevic2011}. Indeed the use of OAM could help improving link capacities as controversially discussed in several recent papers \cite{Tamburini2012}, \cite{Edfors2012}, \cite{Tamagnone2012} \cite{Reply_Tamburini2012}, \cite{CR_Tamagnone2013}, \cite{Tamagnone2015}, \cite{Tamburini2015}.

From the 1990s, the OAM of light has been widely studied in optics with Laguerre-Gaussian beams \cite{Allen1992}, \cite{Padgett1995}, \cite{Gibson2004}, \cite{Twisted_Photons}. The results have been transposed to the radiofrequency domain both theoretically \cite{Thide2007}, \cite{Mohammadi2010} and experimentally \cite{Tamburini2012}. The generation of radio OAM waves can be performed in many ways, \textit{e.g.} with a circular antenna array \cite{Mohammadi2010}, a plane or spiral phase plate \cite{Bennis2013}, \cite{Beijersbergen1994}, or a helicoidal parabolic antenna with dedicated modifications \cite{Tamburini2012}, \cite{Tamburini2015}. The detection appears to be correctly achieved by an interferometer \cite{Tamburini2012} or a 3D vector antenna \cite{Diallo2014}. Such configurations can also be analysed by means of classical communication tools for multiple-input-multiple-output (MIMO) antenna systems \cite{Edfors2012}, \cite{Tamagnone2012}, \cite{Tamagnone2015}.

In \cite{linkbudget2014}, we have analysed the link between two antennas designed to transmit and receive OAM. We have found the same far-zone decay as previously exposed in \cite{CR_Tamagnone2013}. We have also discussed the efficiency of OAM for radio communication. As in \cite{Tamburini2012}, \cite{Edfors2012}, \cite{Tamagnone2012}, \cite{Thide2007} and \cite{Bai2014}, a basic configuration using circular antenna arrays both to transmit and receive one single OAM mode has been studied and will be reused in the present paper. It exhibits the key advantage of a simple model through including the fundamental physical properties of OAM links. 

The wave generated by the system is described by the modal orthogonality of the OAM associated with the rotating helical phase fronts. The antenna elements are
fed with the same signal but at each element of the circular array, their phase vary successively from $0$ to $2\pi l$ circularly around the antenna array axis. The integer $l$ is called the topological charge \cite{Soskin1997}, where in quantum mechanics, $l\hbar$ is the OAM of one photon. The OAM topological charge $l$ is not estimated as in \cite{Mohammadi2010}, but the orthogonality property is exploited here for the possibility to superimpose several channels as it was experimentally performed in optics \cite{Krenn2014} and in RF \cite{Tamburini2012}. 

In this paper, we propose to study the particular properties of the OAM link budget with respect to the OAM topological charge. The aim is to develop an asymptotic far-field formulation valid at large distances between two face-to-face circular array antennas. 
We use the equations of the link budget described in \cite{Edfors2012} and we add an asymptotic analysis that yields physical quantities that are suitable for a system design in a radio communication study with OAM.

The paper is organized as follows. In Section II, we summarize the theory of OAM link budget. In section III, an antenna array configuration is presented and studied first using the superposition principle. In section IV we find and validate an original asymptotic formulation of the link budget. In section V, for any value of the OAM order $l$, we define and study equivalent antenna gains and free-space losses. Finally, in section VI, the calculated formulations are validated using the results of the commercial electromagnetic simulation software Feko.

%---------------------------------------------------------------------
\section{Theory}
\label{Theory}
A link budget addresses the efficiency of a communication system. It takes into account for all the gains and losses from the transmitter to the receiver, associated with both antennas and the propagation channel.

\begin{figure}[!t]
\noindent\includegraphics[width=20pc]{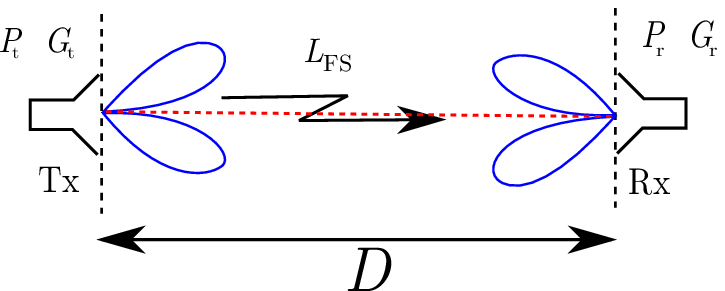}
\caption{Standard configuration for the transmission of waves with non-zero OAM in a single mode configuration. (radiation patterns in blue)}
\label{schemalink}
\end{figure}
In Fig. \ref{schemalink}, $D$ is the distance between the two antennas, $P_{\mbox{t}}$ the transmitted power, $G_{\mbox{t}}$ the transmitter antenna gain, $P_{\mbox{r}}$ the received power, $G_{\mbox{r}}$ the receiver antenna gain and $L_{\mbox{FS}}$ the free-space losses. In its most concise form, $P_{\mbox{r}}$ is given by the Friis transmission equation
\begin{equation}%Pr
P_{\mbox{r}}=\frac{P_{\mbox{t}} G_{\mbox{t}} G_{\mbox{r}}}{L_{\mbox{FS}}}.
\label{Pr}
\end{equation}

In \cite{linkbudget2014}, we have pointed out some difficulties in the calculation of the link budget. Indeed with classical considerations, no power can be received when the antennas are aligned for non-zero OAM orders due to the radiation patterns presented in \cite{Mohammadi2010} and depicted in Fig. \ref{schemalink}. This asymptotic formulation questions the capability of OAM waves to support far-field communications.

%---------------------------------------------------------------------
\section{Link budget for the two circular array configuration} \label{GeneralLinkBudget}
In this section we introduce the equation of the transmission link as proposed by Edfors in \cite{Edfors2012}. In the next section, we go further by determining asymptotic equations that are more suitable for a system design in a radio communication study with OAM.

\subsection{System configuration}
\begin{figure}[!t]
\noindent\includegraphics[width=20pc]{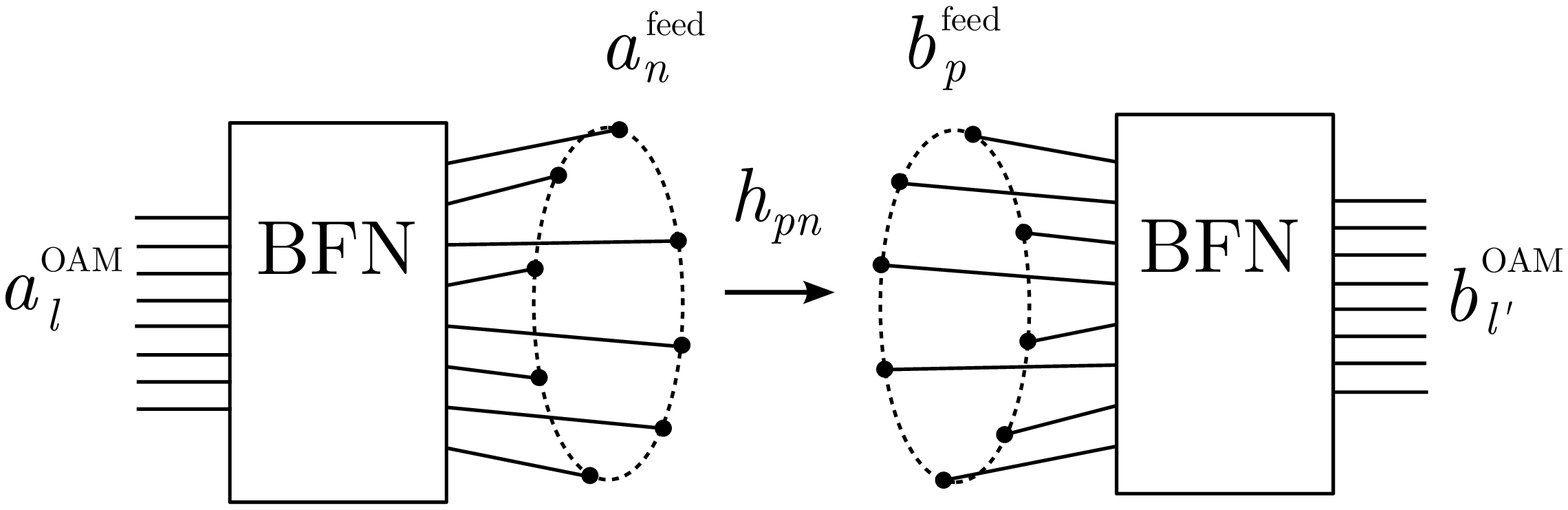}
\caption{Main elements for an OAM transmission with circular arrays.}
\label{BFN}
\end{figure}

In Fig. \ref{BFN} we describe the configuration with which OAM modes of orders $l$ are sent and OAM modes of orders $l'$ are detected.

The transmission can be divided into three blocks: the Beam Forming Network (BFN) of the transmitter, the propagation channel, and the BFN of the receiver. 

In Fig. \ref{BFN}, $a_l^{\mbox{OAM}}$ and $b_{l'}^{\mbox{OAM}}$ are the complex input and output amplitudes of each transmitted or received mode, respectively. Furthermore, $a_n^{\mbox{feed}}$ and $b_{p}^{\mbox{feed}}$ are the wave amplitudes feeding the transmitter array or collected at the receiver array,  respectively. The coefficient $h_{pn}$ corresponds to the propagation term from the element $n$ of the transmitter to the element $p$ of the receiver.

\begin{figure}[!t]
\noindent\includegraphics[width=20pc]{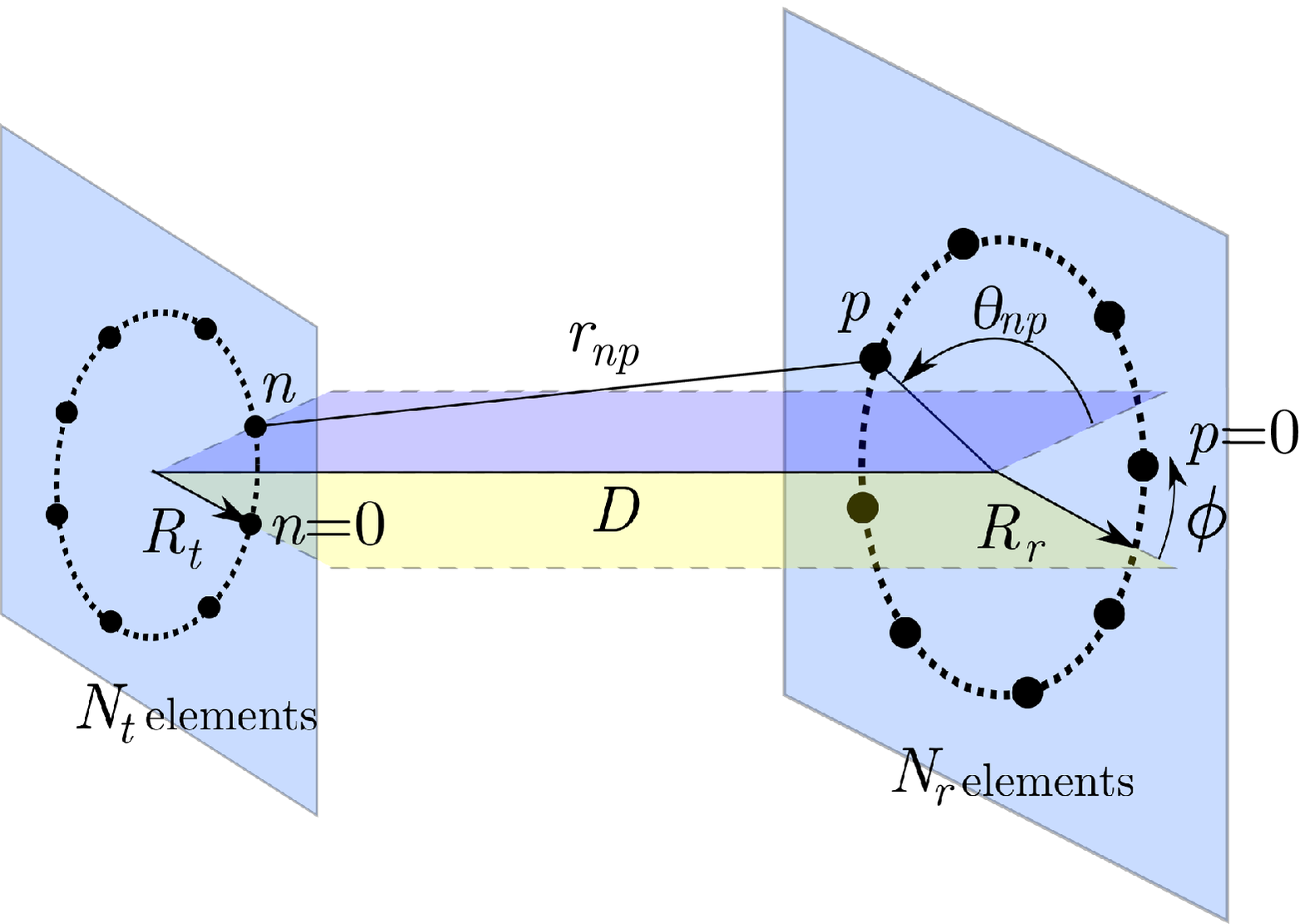}
\caption{Geometry of the circular arrays where the position of each antenna element is equidistant along the perimeters of the circular arrays.}
\label{setup}
\end{figure}

A non-asymptotic configuration is considered in Fig. \ref{setup} as in \cite{Edfors2012} \cite{Mohammadi2010} with all the system parameters depicted.
This system will be thoroughly described afterwards in this section. In \cite{linkbudget2014}, the ratio of the received power of a mode of order $l'$ and the transmitted power carried by a mode of order $l$ has been explicitly formulated and numerically studied.

This article addresses how the link budget expression can be asymptotically expressed in a conventional form as in (\ref{Pr}). For the sake of simplicity, we assume the following hypotheses: 
\begin{enumerate}
\item Each array element is located using its phase center. The polarisation is the same for each antenna element, either linear or circular. 
\item Mutual couplings are neglected.
\item Both BFNs are ideal.
\end{enumerate}
Likewise, some considerations are assumed on the antenna parameters:
\begin{enumerate}
\item The number of antenna elements are the same for both arrays, $N=N_{\mbox{t}}=N_{\mbox{r}}$. $N$ is limited according to the sizes of the antenna array and the antenna elements.
\item The reference tilt angle between the two array antennas is zero, $\phi=0$.
\end{enumerate}
The configuration with two aligned circular array antennas facing each other is depicted in Fig. \ref{setup}. In the following subsections, we study one by one the contribution of the blocks defined in Fig. \ref{BFN}. 

\subsection{BFN Matrix}
On the one hand, to transmit several OAM modes of order $l$ and amplitude $a^{\mbox{OAM}}_{l}$, the antenna elements must be fed by

\begin{equation}%anl
a^{\mbox{feed}}_{n} = \frac{1}{\sqrt{N}} \sum \limits_{l=0}^{N-1} a^{\mbox{OAM}}_{l} e^{- j2\pi \frac{ln}{N}}, 
n\in \left\{0,\ldots,N-1 \right\},
\label{anl}
\end{equation}

with $n$ the element index at the transmitter. This formulation defines an ideal BFN that has $N$ input ports associated with the transmission of OAM modes of order $l \in \left\{0,\ldots,N-1\right\}$. Note that the number of elements in the array limits the number of possible OAM modes due to sampling. Due to aliasing, modes of order greater than $N/2$ are actually modes of negative orders.

On the other hand, the BFN of the receiver builds at its output the OAM modes of order $l'$. The amplitude $b^{\mbox{OAM}}_{l'}$ must be so that

\begin{equation}%bp
b^{\mbox{OAM}}_{l'}=\frac{1}{\sqrt{N}} \sum \limits_{p=0}^{N-1} b^{\mbox{feed}}_{p} e^{ j 2 \pi \frac{pl'}{N} }, 
p\in \left\{0,\ldots,N-1 \right\},
\label{bp}
\end{equation}

with $p$ the element index at the receiver.

From (\ref{anl}), the transmitter BFN matrix that relates the outputs $a^{\mbox{feed}}_{n}$ to the inputs $a^{\mbox{OAM}}_{l}$ is the matrix of the Discrete Fourier Transform (DFT) of size $N$, denoted \textbf{U}. In the same way from (\ref{bp}), the receiver BFN matrix is the inverse discrete Fourier transform and can be characterized by the matrix $\textbf{U}^H$.

\subsection{Channel Matrix}
The propagation channel of this system can be characterized by the channel matrix \textbf{H}. Its terms $h_{pn}$ correspond to the propagation from the phase center of the $n$-th element of the transmitter to the phase center of the $p$-th element of the receiver. The transfer function from the transmitter array and the receiver array is given by \cite{Edfors2012}

\begin{equation}%hpn
h_{pn}= \beta e^{-j k r_{np}} \frac{\lambda}{4\pi r_{np}},
\label{hpn}
\end{equation}

which gives the point-to-point link without coupling terms. The distance between each antenna element is given by

\begin{equation}%
r_{np}= \sqrt{D^2+R^2_{\mbox{t}}+R^2_{\mbox{r}}-2R_{\mbox{t}}R_{\mbox{r}}
\cos\left( \theta_{np} \right)},
\label{rnp}
\end{equation}

with $\theta_{np}=2 \pi \left(\frac{n-p}{N} \right)$.

The free space losses are $4\pi r_{np}/\lambda$, the propagation term is the exponent, $\lambda$ is the wavelength of the carrier.

$\beta$ contains all the variables associated with the antenna system configuration. For the sake of simplicity, the two following hypotheses are added. Firstly, the elements of the transmitter and receiver antennas are in the far-field of each other. Secondly, for large distances between the two arrays, $\beta$ is only related to their gains in the axial direction, therefore it can be approximated by $\sqrt{g_{\mbox{t}} g_{\mbox{r}}}$.

\subsection{Single Mode Link Budget}
Finally, the OAM link can be characterized by the matrix

\begin{equation}%Ht
\textbf{H}_{\textbf{tot}} = \textbf{U}^H \textbf{H} \textbf{U}.
\label{Ht}
\end{equation}

This matrix gives all the relations, \textit{e.g.} the crosstalk, between transmitted and received OAM modes. The output amplitudes are so that

\begin{equation}%bha
\textbf{b}^{\mbox{OAM}}= \textbf{H}_{\textbf{tot}} \textbf{a}^{\mbox{OAM}}.
\label{bHa}
\end{equation}

By expanding the matrix products in (\ref{Ht}) and (\ref{bHa}), the ratio of the received and transmitted powers for only one OAM order $l$ is given by

\begin{equation}%Pr/Pe
\frac{P_{\mbox{r}}}{P_{\mbox{t}}}(l) = \left|  \frac{b^{\mbox{OAM}}_l}{a^{\mbox{OAM}}_l} \right|^2 
= \left|  
\sum \limits_{p=0}^{N-1} \sum \limits_{n=0}^{N-1} \frac{\beta}{N} e^{- j l \theta_{np}}
e^{ -jk r_{np} } \frac{\lambda}{4\pi r_{np}}
\right|^2 .
\label{PrPe2}
\end{equation}

This formulation can be expressed in a different way because the matrix $\textbf{H}$ is circulant. Indeed $h_{p,n}$ only depends on the difference $(n-p)$. From \cite{chan2007introduction} the channel matrix is therefore diagonalized by the $N\times N$ unitary DFT matrix $\textbf{U}$. Besides the eigenvalues $\kappa_l$ of $\textbf{H}$ can be obtained from the DFT of the first row of $\textbf{H}$. This yields

\begin{equation}%
\kappa_l = \sum \limits_{n=0}^{N-1} h_{n,0} e^{-j \frac{2\pi l n}{N} }.
\label{ck}
\end{equation}

Finally, equation (\ref{Ht}) is the diagonalisation of $\textbf{H}$. This means that $\textbf{H}_\textbf{tot}$ is a diagonal matrix of elements $\kappa_l$ and this shows the orthogonality of the OAM modes. The single-mode (\textit{i.e.} $l=l'$) link budget can simply be expressed as 

\begin{equation}%
\frac{P_{\mbox{r}}}{P_{\mbox{t}}}(l) =  \left|\kappa_l\right|^2 = \left| 
\frac{\lambda \beta}{4\pi}
\sum \limits_{n=0}^{N-1}
 \frac{e^{ -jk r_{n0} }}{ r_{n0}}e^{-j 2 \pi  \frac{l n}{N} } 
 \right|^2.
\label{PrPe3}
\end{equation}

This formulation is simpler than (\ref{PrPe2}) for determining the power associated with one OAM mode order.

\section{Asymptotic formulation}
\label{AsympLinkBudget}
\subsection{Development}
The objective of this section is to determine an asymptotic formulation of the link budget (\ref{PrPe3}) at large distances, \textit{i.e.} when $D \rightarrow + \infty$. In other words, we seek the leading term for each value of $l$. To do so, we firstly rewrite (\ref{PrPe3}) using the standard notation $<|>$ for the discrete complex hermitian inner product. This yields

\begin{equation}%
\frac{P_{\mbox{r}}}{P_{\mbox{t}}}(l) = \left| \frac{\lambda \beta}{4\pi} \left\langle u_n \left. \vphantom{\frac{1}{1}} \right| e^{j \frac{2\pi l n}{N} }  \right\rangle \right|^2,
\label{PrPe2sum}
\end{equation}

with $u_n= e^{-jk r_{n0}}/ r_{n0}$.

Since $r_{n0}$ is even with $n$, $u_n$ is even as well. Thus, the previous expression can be written as

\begin{eqnarray}%
\hspace{1.7in} \frac{P_{\mbox{r}}}{P_{\mbox{t}}}(l) & = & \left| \frac{\lambda \beta}{4\pi}  \left\langle u_n \left. \vphantom{\frac{1}{1}} \right|  \cos\left(\frac{2\pi l n}{N}\right)  \right\rangle \right|^2, \nonumber \\
& = & \left| \frac{\lambda \beta}{4\pi} \left\langle u_n \left. \vphantom{\frac{1}{1}} \right|  \cos\left(\frac{2\pi |l| n}{N}\right)  \right\rangle \right|^2.
\label{PrPe4}
\end{eqnarray}

From this expression, the link budget for the modes $+l$ and $-l$ are the same.

To obtain the asymptotic formulation of $P_{\mbox{r}}/P_{\mbox{t}}$, we firstly expand $u_n$ in Taylor series.
For $r_{n0}$, the expression is obtained by writing

\begin{equation}%
r_{n0}= D_{\mbox{tot}} \sqrt{1-\frac{2 R_{\mbox{t}}R_{\mbox{r}}}{D^2_{\mbox{tot}}} \cos\left(\frac{2\pi n}{N}\right)},
\label{rno}
\end{equation}

with $D^2_{\mbox{tot}}=D^2+R^2_{\mbox{t}}+R^2_{\mbox{r}}$. Thus, an asymptotic expansion of $r_{n0}$ when $D_{\mbox{tot}} \rightarrow +\infty$ is given by

\begin{eqnarray}%
\hspace{1in} r_{n0} & = & \sum_{m=0}^{+\infty} \left( \begin{array}{c}
                               1/2 \\ 
                               m 
                              \end{array} \right) \frac{(-2 R_{\mbox{r}} R_{\mbox{t}})^m}{D^{2m-1}_{\mbox{tot}}} \cos^m \left(\frac{2\pi n}{N}\right) \nonumber \\
& = & D_{\mbox{tot}} + \sum_{m=1}^{+\infty} \left( \begin{array}{c}
                               1/2 \\ 
                               m 
                              \end{array} \right) \frac{(-2 R_{\mbox{r}} R_{\mbox{t}})^m}{D^{2m-1}_{\mbox{tot}}} \cos^m \left(\frac{2\pi n}{N}\right),
\label{rno2}
\end{eqnarray}

where $\left( \begin{array}{c}
        \alpha \\ 
        m 
        \end{array} \right)$ 
        are the generalized binomial coefficients.
From this result, we deduce

\begin{eqnarray}%
\hspace{0.7in} e^{-jk r_{n0}} & = & e^{-jk D_{\mbox{tot}}} \nonumber \\
&& \cdot \exp \left[-jk \sum_{m=1}^{+\infty} \left( \begin{array}{c}
                               1/2 \\ 
                               m 
                              \end{array} \right) \frac{(-2 R_{\mbox{r}} R_{\mbox{t}})^m}{D^{2m-1}_{\mbox{tot}}} \cos^m \left(\frac{2\pi n}{N}\right) \right].
\label{ejkrno}
\end{eqnarray}

Similarly, we have 

\begin{equation}%
\frac{1}{r_{n0}} = \frac{1}{D_{\mbox{tot}}} \sum_{m'=0}^{+\infty} \left( \begin{array}{c}
                               -1/2 \\ 
                               m' 
                              \end{array} \right) \frac{(-2 R_{\mbox{r}} R_{\mbox{t}})^{m'}}{D^{2m'}_{\mbox{tot}}} \cos^{m'} \left(\frac{2\pi n}{N}\right). 
\label{1/rno}
\end{equation}

Developing the exponential in (\ref{ejkrno}) and multiplying by (\ref{1/rno}), we finally obtain

\begin{eqnarray}%
\hspace{0.7in} u_n & = & \frac{e^{-j k D_{\mbox{tot}}}}{D_{\mbox{tot}}}\sum_{m'=0}^{+\infty} \left( \begin{array}{c}
                               -1/2 \\ 
                               m'  \end{array} \right)
\frac{(-2 R_{\mbox{r}} R_{\mbox{t}})^{m'}}{D^{2m'}_{\mbox{tot}}} \cos^{m'} \left(\frac{2\pi n}{N}\right) \nonumber \\
&& \cdot \sum_{q=0}^{+\infty} \frac{(-j k)^q}{q!} \left[ \sum_{m=1}^{+\infty} \left( \begin{array}{c}
                                1/2 \\ 
                                m \end{array} \right)
 \frac{(-2 R_{\mbox{r}} R_{\mbox{t}})^m}{D^{2m-1}_{\mbox{tot}}} \cos^m \left(\frac{2\pi n}{N}\right)  \right]^q .
\label{un}
\end{eqnarray}

This expansion is expressed in terms of $\cos^{m} (2 \pi n/N)$ and $\cos^{m'} (2 \pi n/N)$ whereas the initial inner product (\ref{PrPe4}) makes use of $\cos (2\pi |l|n/N)$. Nevertheless, from cosine power reduction formulas and from (\ref{un}), the dominant contribution in $\cos (2\pi |l|n/N)$ is contained in the $\cos^{|l|} (2 \pi n/N)$ term. Indeed, lower-order powers of cosine do not contain terms in $\cos (2 \pi |l| n/N)$ when linearized, and higher-order terms decrease faster when $D\rightarrow \infty$. Therefore, to obtain the link budget (\ref{PrPe4}) for each mode order $l$, the dominant component in $\cos^{|l|} (2 \pi n/N)$ when $D\rightarrow +\infty$ must be extracted. As a consequence of the limit $D\rightarrow +\infty$, the dominant term is obtained only with the components $m=1$ and $m'=0$ in (\ref{un}), and is given by

\begin{equation}
\frac{e^{-j k D_{\mbox{tot}}}}{D_{\mbox{tot}}} \frac{1}{|l|!2^{|l|}} \frac{ (j k 2 R_{\mbox{r}} R_{\mbox{t}})^{|l|}}{D^{|l|}_{\mbox{tot}}} \cos^{|l|}  \left(\frac{2\pi n}{N}\right).
\label{eq_unfin}
\end{equation}

Adding any other terms in the series only yields components decaying faster.

Finally, from (\ref{PrPe4}) and (\ref{eq_unfin}), the link budget is asymptotically given by

\begin{eqnarray}%
\hspace{0.7in} \frac{P_{\mbox{r}}}{P_{\mbox{t}}} (|l|) 
& = &  \left| \frac{\lambda \beta}{4\pi |l|!} \frac{ ( k R_{\mbox{r}} R_{\mbox{t}})^{|l|}}{D_{\mbox{tot}}^{|l|+1}}  \left< \cos^{|l|}  \left(\frac{2\pi n}{N}\right)  \left.  \vphantom{\frac{1}{1}} \right| \cos  \left(\frac{2\pi |l| n}{N}\right)  \right> \right|^2   \nonumber\\
 & = &  \left| \frac{\lambda \beta}{4\pi |l|!}  \frac{ ( k R_{\mbox{r}} R_{\mbox{t}})^{|l|}}{D_{\mbox{tot}}^{|l|+1}} \frac{N}{2^{|l|}} \right|^2.
\end{eqnarray}

Hence, with $D_{\mbox{tot}}\simeq D$ when $D\rightarrow +\infty$, the asymptotic OAM single-mode link budget can be expressed as

\begin{equation}%
\frac{P_{\mbox{r}}}{P_{\mbox{t}}}(|l|) = \left| \frac{\lambda N \beta}{4\pi |l|!} \left( \frac{k R_{\mbox{t}} R_{\mbox{r}}}{2} \right)^{|l|} \frac{1}{D^{|l|+1} }\right|^2.
\label{eq_PrPecalc}
\end{equation}

This result demonstrates that $P_{\mbox{r}}/P_{\mbox{t}}$ decays in $1/D^{2|l|+2}$ for a mode of order $|l|$.

\subsection{Validation and convergence}

\subsubsection{Convergence}
In this section we want to study the convergence of the global link budget (\ref{PrPe3}) with the asymptotic formulation (\ref{eq_PrPecalc}). In addition, the dependence on the OAM order $l$ is also studied.

The difference between the asymptotic formulation (\ref{eq_PrPecalc}) and the non-asymptotic one (\ref{PrPe3}) can be estimated. 
For any distance, we can compute the relative difference given by

\begin{equation}%
\mathcal{E} = \frac{|P_{\mbox{r}}/P_{\mbox{t}} -(P_{\mbox{r}}/P_{\mbox{t}})_{\mbox{asymp}}|}{|P_{\mbox{r}}/P_{\mbox{t}}|}.
\label{eq_diff}
\end{equation}

\begin{figure}[!t]
\noindent\includegraphics[width=20pc]{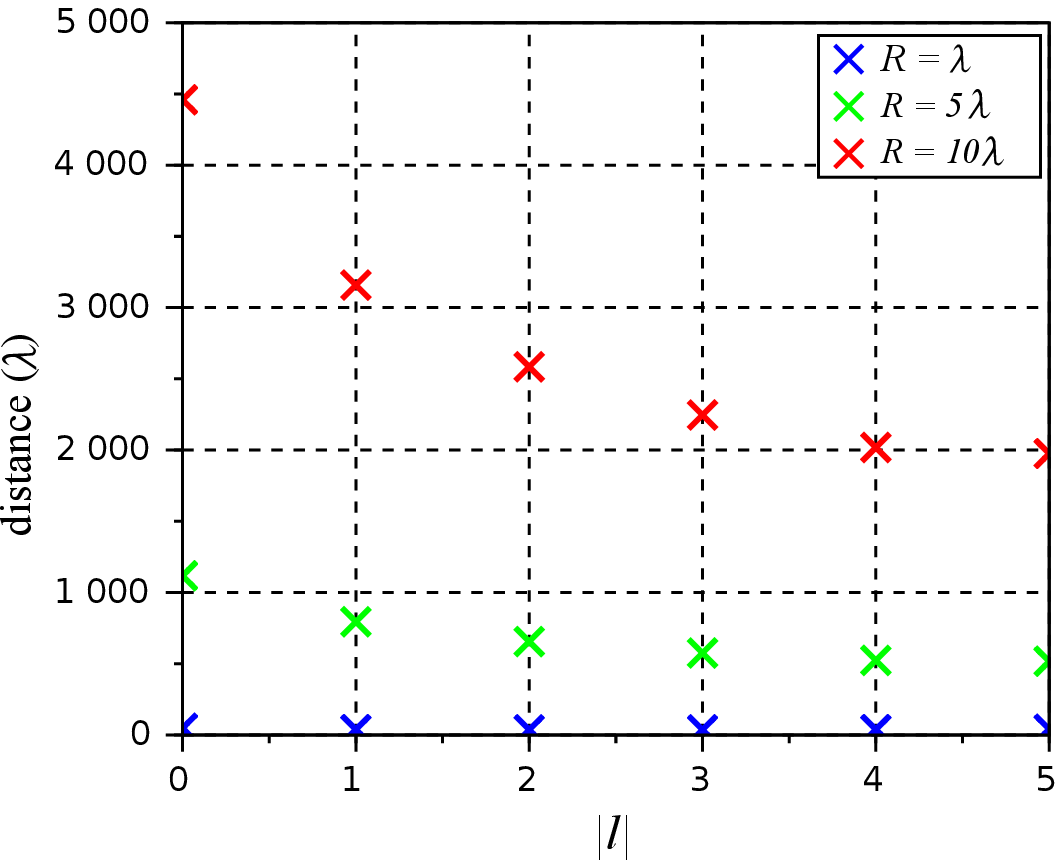}
\caption{Distance from which $1\%$ relative difference level is reached with $N=12$ and for different values of $R=R_{\mbox{t}}=R_{\mbox{r}}$.} %légende
\label{farfielddistance} %label
\end{figure}

From the results in Fig. \ref{farfielddistance}, the distances at which the relative difference levels are at least of $1\%$ depend on $l$ but the farthest distance is given by the OAM mode order $l=0$. Therefore the traditional method to determine the far field distance remains the same and can be applied for every OAM mode order and are conventionally dependent on the radii. Thus, in the far field, the asymptotic formulation (\ref{eq_PrPecalc}) holds.

\subsubsection{Asymptotic slope}
\label{slope}
\begin{figure}[!t]
\noindent\includegraphics[width=20pc]{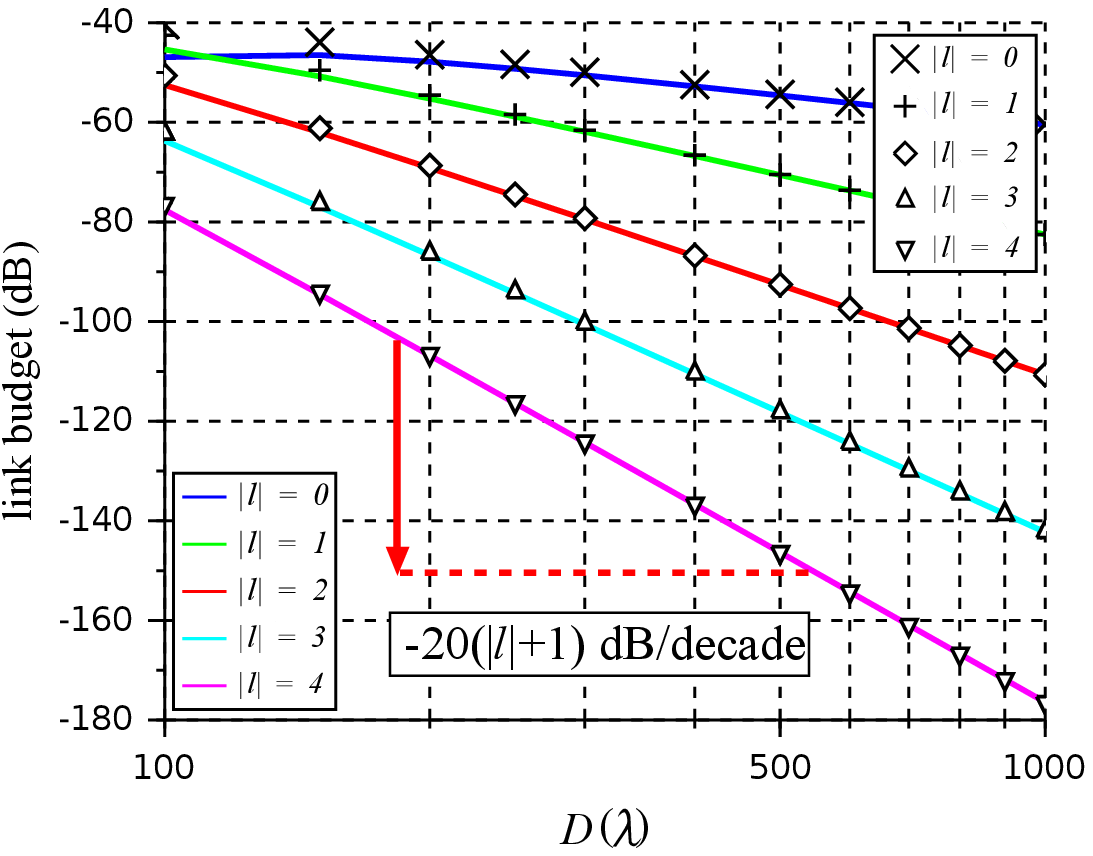}
\caption{Link budgets for $|l|=0,1,2,3,4$ with $N=12$, $R=R_{\mbox{t}}=R_{\mbox{r}}=5 \lambda$ and $g_{\mbox{t}}=g_{\mbox{r}}=1$. The link budget (\ref{PrPe3}) is depicted in color lines, and the asymptotic formulation (\ref{eq_PrPecalc}) is represented with dots.} %légende
\label{fig_bilan} %label
\end{figure}

In order to study the formulation (\ref{eq_PrPecalc}), we show in Fig. \ref{fig_bilan} the link budgets from (\ref{PrPe3}) and (\ref{eq_PrPecalc}) computed for $|l|=0,1,2,3,4$ and $R_{\mbox{t}}=R_{\mbox{r}}=5\lambda$. As expected, the two representations converge.
From the Fraunhofer distance $2(2 \,\mbox{max}(R_{\mbox{t}},R_{\mbox{r}}))^2/\lambda= 200 \lambda$, the link budget asymptotically tends to straight lines of slope $-20(|l|+1)$ dB/decade, which is consistent with an attenuation in $1/D^{2|l|+2}$. This was first discovered in \cite{Tamagnone2012} for the case $l=1$ with a two element antenna array system, and extended for all the OAM mode orders in \cite{CR_Tamagnone2013} with the Laguerre-Gaussian beam formalism. Here we confirm this result and extend it for any circular array antenna system. This result highlights the strong asymptotic slope for non-zero OAM mode orders. This will be discussed in the following sections and in the conclusion.

Finally, with this expression (\ref{eq_PrPecalc}), we can calculate immediately the link budget. In addition, we have validated the decay in $1/D^{2|l|+2}$ of waves carrying OAM. In the next section we study the asymptotic formulation of the link budget (\ref{eq_PrPecalc}) in terms of gains and losses.

%--------------------------------------------------------------------------
\section{Gain and free-space losses}
\subsection{Asymptotic Expressions}
In this section, we generalise the classical transmission equation (\ref{Pr}) to non-zero OAM mode order. To do so, we define equivalent antenna gains $G_{\mbox{t}_{\mbox{eq}}}$, $G_{\mbox{r}_{\mbox{eq}}}$ and equivalent losses $L_{\mbox{FS}_{\mbox{eq}}}$. They depend on the characteristics of the antennas, on the distance and on the OAM order $l$.

Keeping in mind $L_{\mbox{FS}} = \left( 4\pi D/\lambda \right)^{2}$, we want to find a formulation with only gains and losses, \textit{i.e.}

\begin{equation}%
\frac{P_{\mbox{r}}}{P_{\mbox{t}}}(l) = G_{\mbox{t}_{\mbox{eq}}}(l) \, G_{\mbox{r}_{\mbox{eq}}}(l) \, \frac{1}{L_{\mbox{FS}_{\mbox{eq}}}(l)}.
\label{PrPeAsymp}
\end{equation}

With some transformations in (\ref{eq_PrPecalc}), we can obtain

\begin{equation}%
 \frac{P_{\mbox{r}}}{P_{\mbox{t}}}(l)  = \frac{N g_{\mbox{t}}}{|l|!}  \left( \frac{4\pi (\pi R^2_{\mbox{t}})}{\lambda^2} \right)^{|l|}  \frac{N g_{\mbox{r}}}{|l|!}  \left( \frac{ 4\pi (\pi R^2_{\mbox{r}})}{\lambda^2} \right)^{|l|}  \left( \frac{\lambda}{4\pi D} \right)^{2|l|+2} .
\label{PrPeanalog}
\end{equation}

In this equation, we can identify equivalent free-space losses

\begin{equation}%
L_{\mbox{FS}_{\mbox{eq}}}(l) = \left( \frac{4\pi D}{\lambda} \right)^{2|l|+2}.
\label{LFSl}
\end{equation}

Note that the free-space losses have exactly the classical expression $L_{\mbox{FS}}=\left( 4\pi D/\lambda \right)^2$ but with an exponent $|l|+1$, which gives $L_{\mbox{FS}_{\mbox{eq}}}(l) = (L_{\mbox{FS}})^{|l|+1}$.

From (\ref{PrPeAsymp}) and (\ref{PrPeanalog}), we also define equivalent OAM antenna gains for a configuration with two face-to-face circular array antennas. Their expressions are given by

\begin{equation}%
G_{\mbox{eq}}(l) = \frac{N g}{|l|!} \left( \frac{4\pi (\pi R^2)}{\lambda^2} \right)^{|l|}.
\label{Gl}
\end{equation}

Since from the beginning we have separated the transmitter and receiver parts, the formulation (\ref{PrPeanalog}) remains valid for a communication system with two different array antenna radii. In addition, this formula is only valid for array antennas.

\subsection{Equivalent array antenna gain study}
\label{EqGain}
The equivalent gain (\ref{Gl}) is dependent on the antenna parameters. 
When $l=0$, we have $G_{\mbox{t}_{\mbox{eq}}}=G_{\mbox{r}_{\mbox{eq}}}=Ng$ which is classical for an array antenna with elements fed in-phase with neglected couplings. When $l\neq 0$, $G_{\mbox{t}_{\mbox{eq}}}$ and $G_{\mbox{r}_{\mbox{eq}}}$ are dependent on $N$, $R_{\mbox{t}}$, $R_{\mbox{r}}$ and $l$. Therefore, the transmitted and received equivalent gains can be modified to partially compensate the free-space losses in $1/D^{2|l|+2}$ at a given distance.
\begin{figure}[!t]
\noindent\includegraphics[width=20pc]{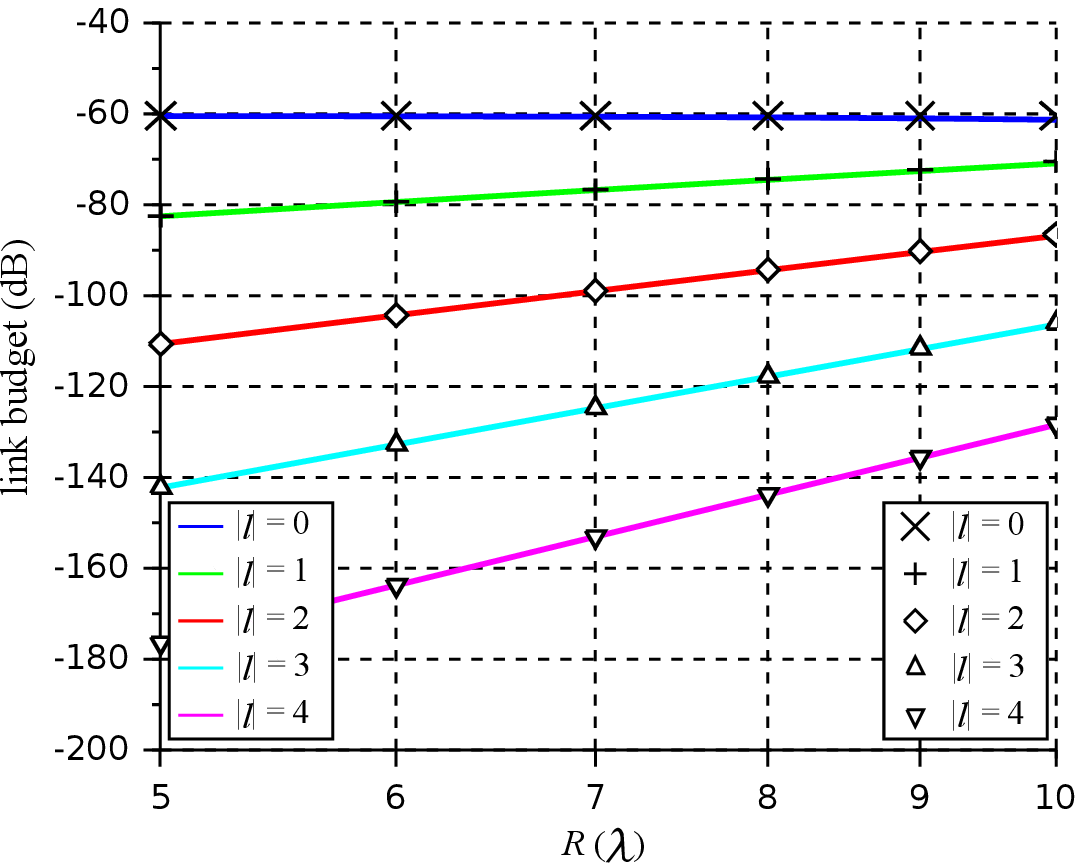}
\caption{Link budgets for $|l|=0,1,2,3,4$ with $N=12$, $R=R_{\mbox{t}}=R_{\mbox{r}}$, $g_{\mbox{t}}=g_{\mbox{r}}=1$ and $D=1000 \lambda$. The link budget (\ref{PrPe3}) is depicted in color lines, and the asymptotic formulation (\ref{eq_PrPecalc}) is represented with dots.} %légende
\label{fig_bilan_R} %label
\end{figure}
In Fig. \ref{fig_bilan_R}, equation (\ref{PrPeanalog}) of the link budget is computed for a transmission between two circular array antennas separated by a distance of $D=1000 \lambda$ where $R=R_{\mbox{t}}=R_{\mbox{r}}$ varying from $1 \lambda$ to $10 \lambda$.
A comparison with (\ref{PrPe3}) is also plotted to confirm the asymptotic behaviour.
For a fixed value of $|l|$, each equivalent gain increases in $R^{2|l|}$ so that the link budget improves by a factor of $R^{4|l|}$. On the contrary, for a fixed value of $R$, when $|l|$ increases, the link budget decreases since asymptotically the effect of $D$ is greater than those of $R_{\mbox{t}}$ and $R_{\mbox{r}}$.

%------------------------------------------------------------------------------
\section{Simulation}
A general asymptotic formulation (\ref{PrPeanalog}) of the link budget has been obtained and validated. It matches the classical theory (\ref{Pr}) and can give the OAM single-mode link budget. In this section, another validation with the commercial electromagnetic simulation software Feko is performed.

\subsection{Setup}
The same system as before with two identical face-to-face array antennas is modelled in the software. We choose to have 8 elements per array. This will allow the generation of the OAM mode orders $|l|=0,1,2$. 

\subsubsection{Array antenna element}
\begin{figure}[!t]
\noindent\includegraphics[width=20pc]{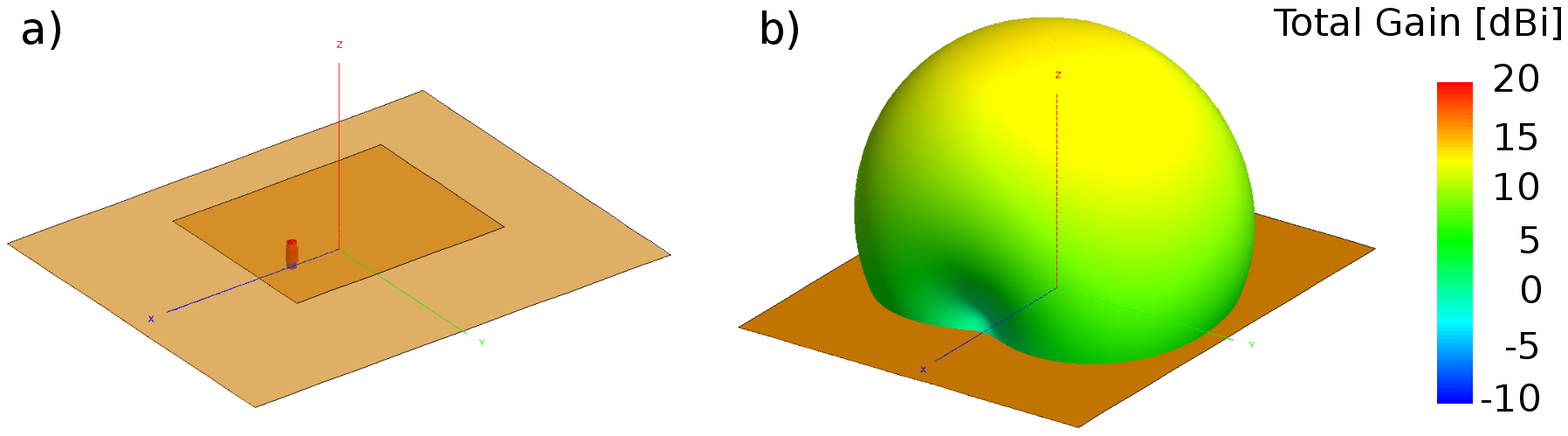}
\caption{(a) 3D model of a single rectangular patch designed for $2.42$GHz and only one linear polarisation. (b) Its radiation pattern.  } %légende
\label{Feko_patch} %label
\end{figure}

For the array elements, we use rectangular patch antennas designed to work at $2.42$GHz, as depicted in Fig. \ref{Feko_patch}. For the sake of simplicity, the dielectric substrate is air and the patch is fed by a wire source between the patch and the ground plane in such a way that a linear polarisation is excited.
In the simulation results, the antenna gain of the element is $8.70$dBi. This value is given to the terms $g_{\mbox{t}}$ and $g_{\mbox{r}}$ in the asymptotic formulation (\ref{PrPeanalog}).

\subsubsection{Array antenna}
\begin{figure}[!t]
\noindent\includegraphics[width=20pc]{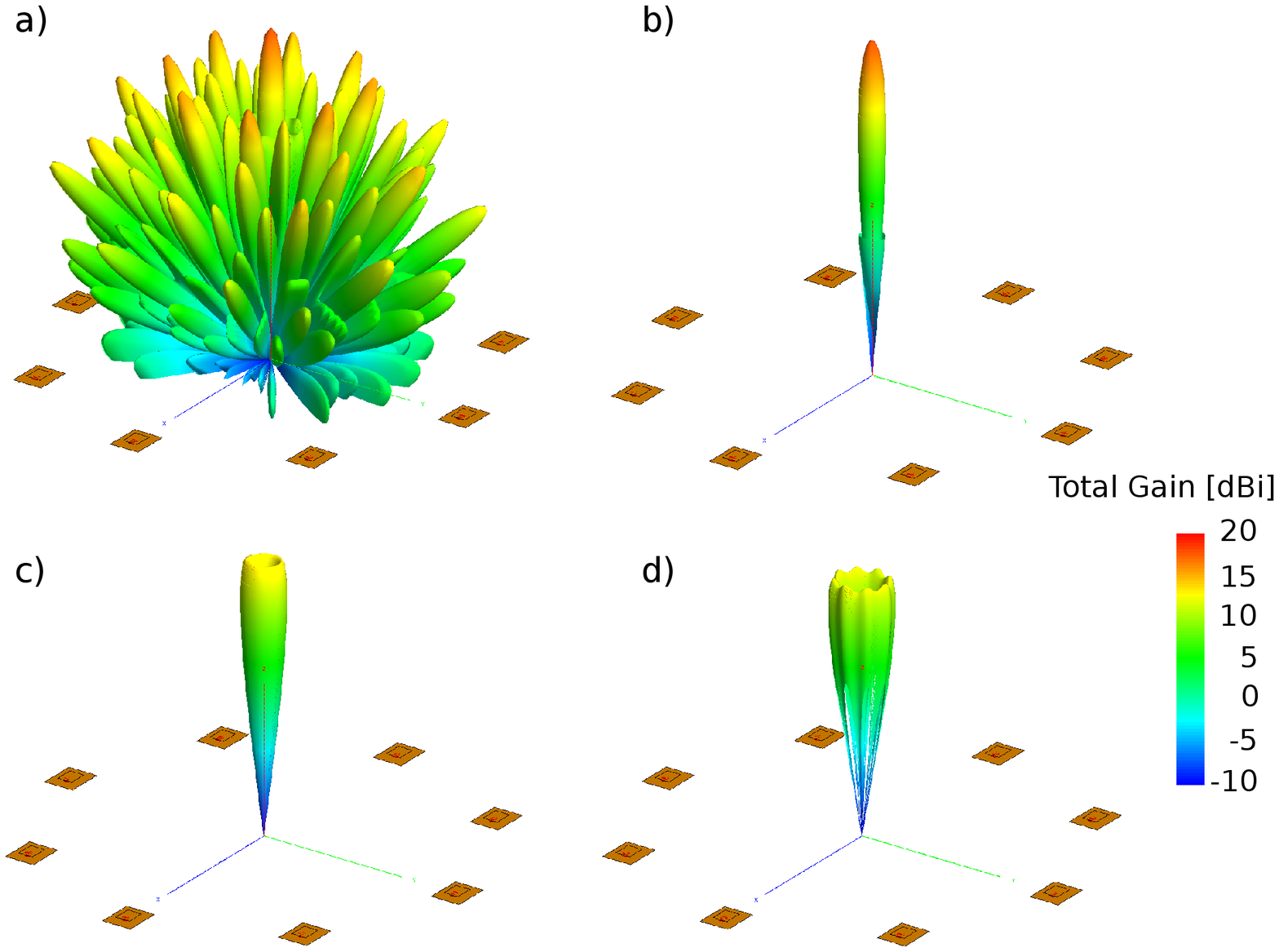}
\caption{(a) Radiation patterns of the circular array for the OAM mode order $l=0$. Due to the high grating lobes (b) (c) (d) show the radiation patterns for the OAM mode orders $|l|=0,1,2$ in the vicinity of the array axis.} %légende
\label{Feko_array} %label
\end{figure}

The 8 element array and its radiation patterns for OAM mode orders $|l|=0,1,2$ are represented in Fig. \ref{Feko_array}. 
There are many grating lobes caused by the annular geometry of the array. However, along the array axis, a good OAM radiation pattern is generated and is convenient for the simulation.

The far field is here obtained beyond the Fraunhofer distance located at $200\lambda$ for the arrays of radii $5 \lambda$. The simulations are performed with two face to face arrays using the same polarisation. 

The S-parameters between each port of the arrays are computed to obtain the simulated propagation channel matrix so that all the couplings are directly included in the S-parameter.
The BFN is still considered as ideal. So the matrix equation (\ref{bHa}) can be used to calculate all the link budgets but now with simulated values for $\textbf{H}_\textbf{tot}$, taking mutual couplings into account.

\subsection{Parametric Simulations}
\subsubsection{Distance}
\begin{figure}[!t]
\noindent\includegraphics[width=20pc]{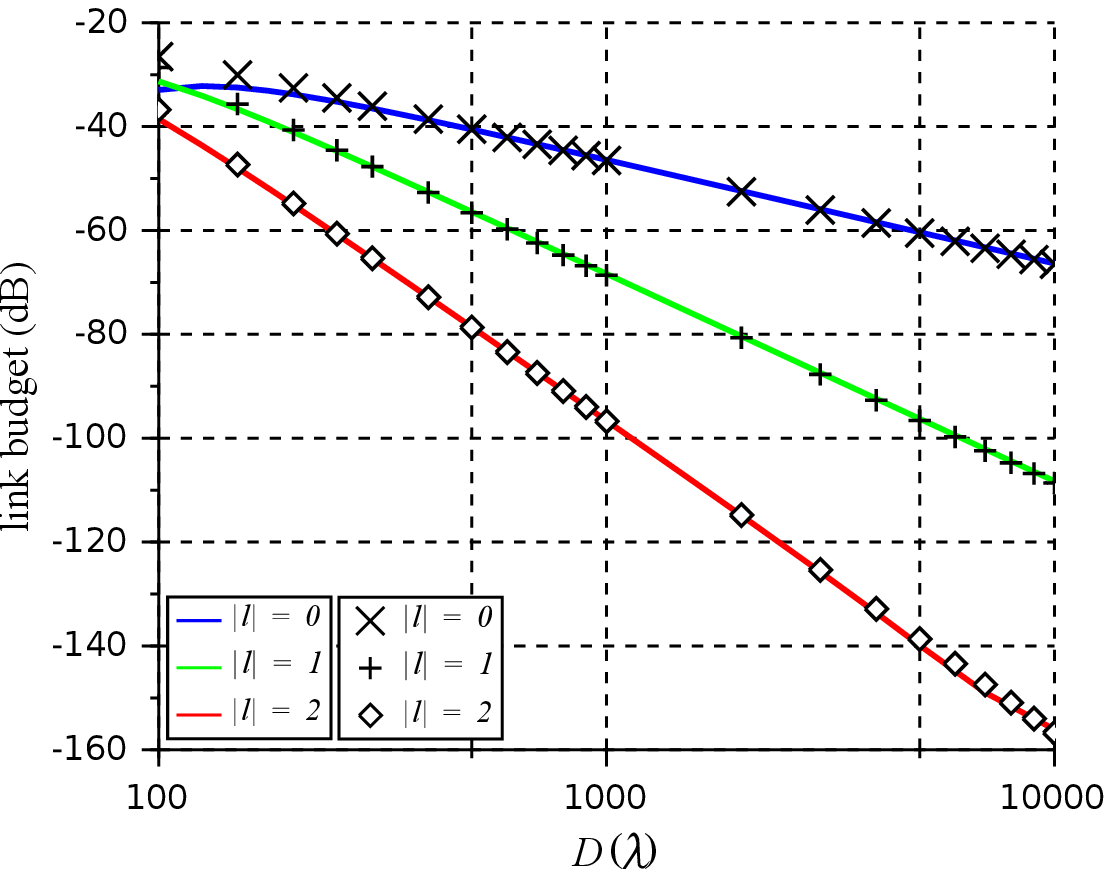}
\caption{Simulated link budget between the two array antennas of radii $5\lambda$ in function of the distance $D$. The simulated link budget is depicted in color lines, and the calculated link budget with the asymptotic formulation (\ref{PrPeanalog}) is represented with dots.} %légende
\label{Feko_D} %label
\end{figure}

First, we want to verify the asymptotic slope in $1/D^{2|l|+2}$. A parametric study is performed on the distance between the two array antennas and
the results are displayed in Fig. \ref{Feko_D}.
The results show a good match and Fig. \ref{Feko_D} clearly shows the slope in $1/D^{2|l|+2}$ beyond the Fraunhofer distance of $200\lambda$. 

\subsubsection{Radii}
\begin{figure}[!t]
\noindent\includegraphics[width=20pc]{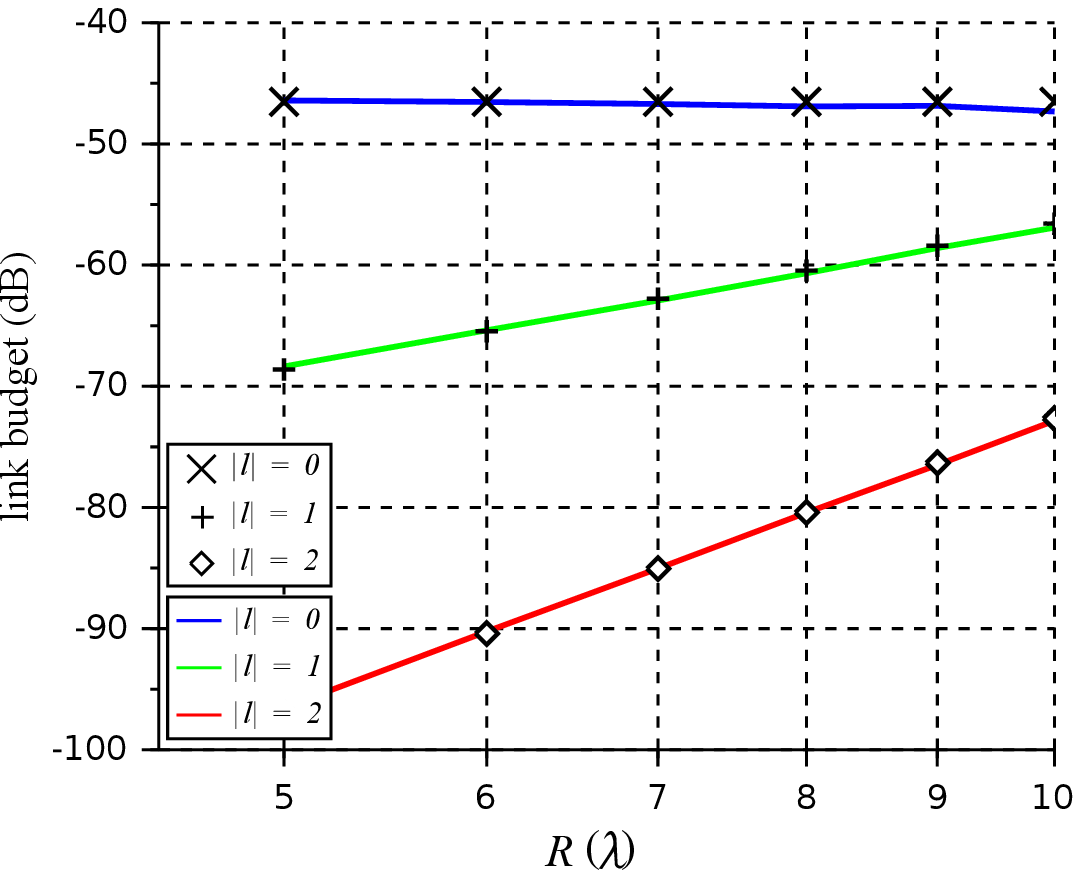}
\caption{Simulated link budget between both array antennas separated by $1000\lambda$ in function of the array radii. The simulated link budget is depicted in color lines, and the calculated link budget with the asymptotic formulation (\ref{PrPeanalog}) is represented with dots. } %légende
\label{Feko_R} %label
\end{figure}

The antenna gain behaviour with respect to the array radii, highlighted in section \ref{EqGain}, is verified in the simulation depicted in Fig. \ref{Feko_R}.
The results show a good match and illustrate the usefulness of changing the array radii to optimise the link budget.

\subsection{Conclusion on the Simulations}
In this section, we have compared the simulated link budget with the asymptotic formulation (\ref{PrPeanalog}). The results show a very good match and validate the formulations of Section \ref{AsympLinkBudget}. 

Finally, for circular arrays, the asymptotic formulation (\ref{PrPeanalog}) allows to calculate the link budget of an OAM system instantly. As for the classical Friis transmission equation, the asymptotic formulation (\ref{PrPeanalog}) can be very useful in a OAM system pre-design and clearly shows the impact of each parameter of the system on the link budget. The good match in the results confirms that the common hypotheses taken beforehand, like neglecting couplings, can be consistent in these conditions.

For non-zero OAM mode, the high asymptotic slope due to the free space losses is determining in the system design. A trade-off with the antenna gains can be achieved to reach the requirement. However, as shown in \cite{Edfors2012}, in the very far field, the higher order OAM waves become rapidly weak and only the $l=0$ OAM mode remains usable.

\section{Conclusion}

We have presented an original formulation of the link budget with equivalent OAM gains and free-space losses dependent on the OAM order $l$. For an OAM system design, it clearly shows the impact of each system parameter on the link budget.

In this paper, we have investigated the asymptotic formulation of the OAM link budget. From the classical theory we have seen that the OAM link budget cannot be calculated as usual. We have also studied a configuration with two circular array antennas. This system is simple but sufficient enough to study the general properties of the OAM. We have found an asymptotic formulation of the link budget by means of asymptotic expansions. An asymptotic decay in $1/D^{2|l|+2}$ has been observed. The calculation have shown that the asymptotic formulation holds from a distance which classically depend on the OAM mode order $l=0$. The equivalent gain formula has been studied and highlights the influence of $R$, $l$ and of the array elements on the link budget. 
Finally, the calculated formulations have been validated with the results of the commercial electromagnetic simulation software Feko. This clearly shows the advantage of our formulas for a rapid system design.

The study has shown some elements on the link budget for the use of the new degree of freedom offered by the OAM in the case of superimposition of several channels on the same bandwidth. The proposed formulation is specific to face-to-face circular arrays but the theory may be extended to generalize this formulation to other configurations, \textit{e.g.} a continuous aperture. 

Nonetheless, the results that we have shown confirm the difficulties to use OAM at very large distance. However, at shorter distances, they remain of interest because there exists communication configurations in which the link budget is more suitable.

%%% End of body of article:

%%%%%%%%%%%%%%%%%%%%%%%%%%%%%%%%
%% Optional Appendix goes here
%
% \appendix resets counters and redefines section heads
% but doesn't print anything.
% After typing \appendix
%
%\section{Here Is Appendix Title}
% will show
% Appendix A: Here Is Appendix Title
%
%%%%%%%%%%%%%%%%%%%%%%%%%%%%%%%%%%%%%%%%%%%%%%%%%%%%%%%%%%%%%%%%
%
% Optional Glossary or Notation section, goes here
%
%%%%%%%%%%%%%%
% Glossary is only allowed in Reviews of Geophysics
% \section*{Glossary}
% \paragraph{Term}
% Term Definition here
%
%%%%%%%%%%%%%%
% Notation -- End each entry with a period.
% \begin{notation}
% Term & definition.\\
% Second term & second definition.\\
% \end{notation}
%%%%%%%%%%%%%%%%%%%%%%%%%%%%%%%%%%%%%%%%%%%%%%%%%%%%%%%%%%%%%%%%
%
%  ACKNOWLEDGMENTS

\begin{acknowledgments}
We thank Direction G\'en\'erale de l'Armement (DGA) and Centre National d'Etudes Spatiales (CNES) for their financial support to this work. Data obtained in the paper can be fully recalculated through any numerical computational tool with the formulations presented in the paper. Simulated data are calculated with the commercial electromagnetic simulation software Feko.
\end{acknowledgments}

\end{article}
%
%
%% Enter Figures and Tables here:
%
% DO NOT USE \psfrag or \subfigure commands.
%
% Figure captions go below the figure.
% Table titles go above tables; all other caption information
%  should be placed in footnotes below the table.
%
%----------------
% EXAMPLE FIGURE
%
 %\begin{figure}
 %\noindent\includegraphics[width=20pc]{samplefigure.eps}
 %\caption{Caption text here}
 %\label{figure_label}
 %\end{figure}
%
% ---------------
% EXAMPLE TABLE
%
%\begin{table}
%\caption{Time of the Transition Between Phase 1 and Phase 2\tablenotemark{a}}
%\centering
%\begin{tabular}{l c}
%\hline
% Run  & Time (min)  \\
%\hline
%  $l1$  & 260   \\
%  $l2$  & 300   \\
%  $l3$  & 340   \\
%  $h1$  & 270   \\
%  $h2$  & 250   \\
%  $h3$  & 380   \\
%  $r1$  & 370   \\
%  $r2$  & 390   \\
%\hline
%\end{tabular}
%\tablenotetext{a}{Footnote text here.}
%\end{table}

% See below for how to make sideways figures or tables.

\end{document}